# COLLINEAR TWO COLOUR KERR EFFECT BASED TIME-GATE FOR BALLISTIC IMAGING

Harsh PURWAR*, Saïd IDLAHCEN, Claude ROZÉ and Jean-Bernard BLAISOT

CORIA-UMR 6614-Normandie Université, CNRS-Université et INSA de Rouen, Campus Universitaire du Madrillet, 76800 Saint Etienne du Rouvray, France

*Corresponding author: harsh.purwar@coria.fr

**Abstract**

A novel setup is presented for ballistic imaging using an efficient ultrafast Kerr effect based optical time-gate with gating times of the order of ~0.8 picoseconds. At first, the major drawbacks of the classical non-collinear optical setup are discussed. Then, the new collinear arrangement is proposed, which overcomes these issues and improves the achievable imaging spatial resolution and gate timings. Few preliminary results for ballistic imaging of liquid sprays/jets are presented for this arrangement. It is shown that using a solid state Kerr medium (GGG crystal), instead of the classical liquid $CS_2$, allows reduction in the opening time of the optical gate.

## 1 Introduction

Over the past few decades there has been a tremendous effort in developing optical diagnostic tools, mainly due to their non-intrusive nature, for application in the various fields of science and technology, like for imaging of biological tissues [1-3], measurements of high-pressure multiphase flows [4,5] etc. However, a common issue experienced in general, with these diagnostic tools is due to the fact that most of the real world applications are linked to a turbid environment, which largely restricts the possibilities of these tools. The key information is scrambled by the distortion imparted to the light signal when it transits the measurement volume. Informative optical diagnostics in such media require detailed understanding of the light source, propagation and scattering in the measurement volume, and a detection arrangement tailored to select the meaningful parts of the transmitted light signal.

To this end, ultrafast time gating can provide an effective means of segregating high integrity portions of the collected signal from light disturbed by scattering interactions. On average, photons that participate in more interaction events traverse a more circuitous path through the medium and are statistically more likely to be distorted or redirected from their original trajectories. This difference in optical path length results in a temporal spreading of the light intensity such that heavily distorted signal components arrive at later times. Time gating, or time filtering allows selection of the optical signal within a small temporal window and could be useful in separating the signal that is scrambled due to multiple reflections or scattering from the signal that retains the interpretable information on the object characteristics [6].

Nevertheless, the time resolution of the time gating must be short: assuming a 100 fs short input pulse and a ~1 cm turbid measurement volume, the typical transmitted pulse durations can be expected to be of the order of 50-100 ps, with most of the informative signal present in the first 500 fs of the transmitted signal [4]. Consequently, most time-gating applications require arrangements that can limit light collection to a 1 ps or shorter time window.

## 2 Non-collinear, single color configuration

A schematic of the optical setup for ultrafast Kerr effect based time gate with the non-collinear incidence of the pump and probe beams and $CS_2$ as the Kerr medium is shown in Figure. 1.

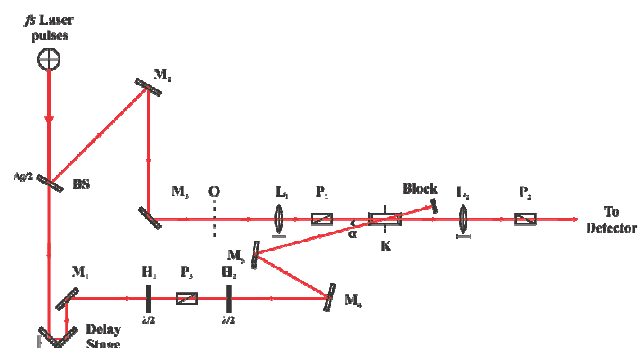

*Figure 1* Schematic of the experimental setup for the classical OKE-based time-gate with non-collinear incidence of the pump and probe beams at the Kerr medium ($CS_2$ in this case). BS: 50/50 beam splitter, ($M_1$, $M_2$, $M_3$, $M_4$, $M_5$): mirrors, ($P_1$, $P_2$, $P_3$): polarizers, ($H_1$, $H_2$): half waveplates, O: object plane, ($L_1$, $L_2$): bi-convex lenses with focal length f = 200 mm, α: angle between the pump and probe beams.

The light source consists of femtosecond laser pulses generated using a regenerative amplifier (Libra, Coherent) seeded with a Titanium-Sapphire mode-locked laser (Vitesse, Coherent). The system is capable of generating laser pulses with pulse width less than 120 fs and energy of about 3.5 mJ per pulse at a repetition rate of 1 kHz, pulse spectrum centred at wavelength λ = 800 nm. The incoming fs laser beam was separated into probe and pump beams using a 50:50 beam splitter (BS). The pump pulse, used to activate Kerr effect in the Kerr medium (K) for a short duration, depending on the non-linear properties of the medium and on the pump pulse characteristics, passes through a computer controlled delay stage with a least count of 0.1 μm (corresponding to a temporal delay of 0.67 fs in air). The delay between the pump and the probe pulses is adjusted so that both pulses arrive at the Kerr medium at the same time and a good alignment takes care of their spatial overlap. A combination of a half waveplate ($H_1$) and a polarizer ($P_3$) is used to adjust the power of the pump beam and another half waveplate ($H_2$) to rotate its polarization axis by an angle, with respect to the probe polarization, which maximizes the induced birefringence in the Kerr medium, thus maximizing the efficiency of the time-gate.

The probe pulse illuminates the object (O) under study and is extended in time due to the interaction with the object. It then passes through the Kerr medium, which is sandwiched between two crossed polarizers $P_1$ and $P_2$. The pump, i.e. the switching pulse, induces a transient, intensity dependent birefringence in the Kerr medium, which changes the polarization of a part of the probe pulse, which overlaps with the pump both spatially and temporally and allows it to pass through the second polarizer $P_2$. An additional neutral density filter may be used to adjust the intensity of light reaching the detector to comply with its limitations. For imaging purposes a CCD Camera (Hamamatsu C9100-02) was used as a detector whereas for basic intensity measurements a power meter (Ophir Nova-II) was used.

Major issues in this classical non-collinear OKE-based time gating configuration arise due to the angle between the pump and probe beams. Both gate duration and efficiency are affected due to this. Also, since the overlap of these beams inside the Kerr medium is non-collinear, the gate characteristics also depend on the physical sizes of these beams inside it.

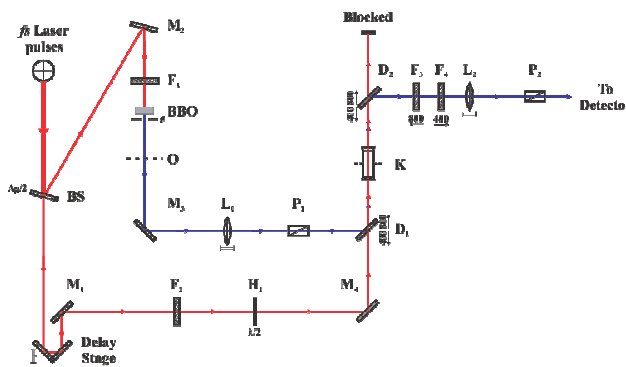

*Figure 3* Schematic of the experimental setup for dual colour OKE-based time-gate with collinear incidence of the pump and probe beams at the Kerr medium. BBO: β-barium borate crystal, ($D_1$, $D_2$): dichroic mirrors, ($F_1$, $F_2$): neutral density filters, ($F_3$, $F_4$): edge-pass and band-pass filters respectively to block pump beam (rest of the symbols have the same meaning as in Figure 1).

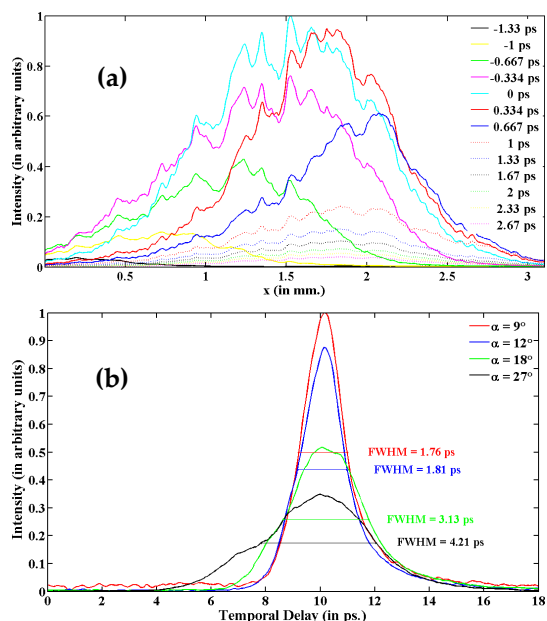

*Figure 2* (a) Average line profiles of the time-filtered images (without any object/sample) and (b) angular dependence of the gate characteristics for single colour, non-collinear configuration with 1.0 mm liquid $CS_2$ as Kerr medium.

To illustrate this issue we show the average line profiles for the intensity images obtained for different delays between the pump and probe pulses when no object (O) is inserted in the probe's path in Figure 2(a). The temporal profile dependence of the optical gate on the angle between the pump and probe beams is shown in Figure 2(b). When sub-picosecond gate time resolution is targeted, the delay between the two sides of the image is a real drawback of this setup.

It is clear from these results that the angle between the pump and probe pulses must be reduced to a minimum to obtain an ultrafast optical time gate. But as one reduces the angle between these pulses, at first, their separation after the interaction inside Kerr medium becomes difficult and then the noise due to the scattering of the high power pump beam from $CS_2$ affects the obtained results. To overcome these issues we propose the following configuration for the optical time-gate.

## 3 Collinear, dual colour configuration

A major difference in this setup compared to other existing setups in literature [7-9] is that here the pump and probe beams are incident at the Kerr medium in a collinear fashion, which resolves most of the major issues with optical Kerr gating. Here, a β-barium borate (BBO) crystal is introduced in the path of the probe beam before it illuminates the object (O) to change its wavelength (from λ

= 800 nm to 400 nm) using the principle of second harmonic generation. The pump and probe beams now have different wavelengths and a pair of dichroic mirrors is used for combining and separating these pulses such that they interact inside the Kerr medium collinearly.

They are again separated afterwards, so that the probe is only detected (as shown in Figure 3). The spatial overlap of the pump beam on the probe beam covers an area in the Kerr medium corresponding to a constant time. This resolves the problem of mixing of spatial and temporal information carried by the probe as it passes through the optical gate [9]. Moreover, the noise due the scattering of the pump is naturally removed by spectral filtering. The average line profiles for the intensity images obtained for different delays between the pump and the probe beams in this collinear configuration are shown in Figure 4. There is no spatial spread/error and hence the temporally filtered images obtained with this setup contain spatial information pertaining to the same time event.

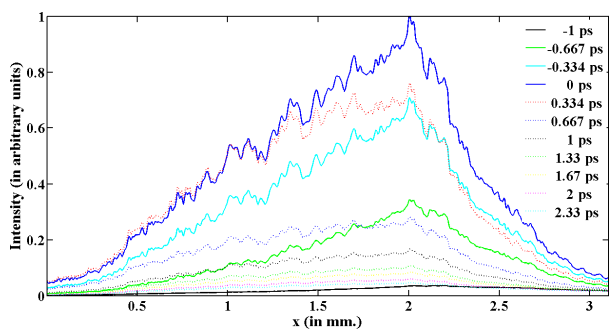

*Figure 4* Average line profiles for the time-filtered images (without any object/sample) with 1.0 mm $CS_2$ as Kerr medium for collinear, dual colour configuration of the optical time-gate.

It should be noted that in this configuration the optical gate characteristics, including the efficiency, does not depend on the physical beam sizes as long as the size of the pump beam is big enough to overlap the probe.

A consequence of using different wavelengths for pump (800 nm) and probe (400 nm) beams which is not present in the classical single colour, non-collinear configuration is the dependence of the optical gate duration on the thickness of the Kerr medium. This is due to the fact that both beams now have different group velocities inside the Kerr medium. So for a suitable delay, even if the pump and probe pulses do not overlap outside the Kerr medium, pump can catch up with the probe and the overlap can still occur inside the medium. However, occurrence of such an event is reduced when the thickness of the Kerr medium is decreased. Figure 5 shows the time profile for two different thicknesses of the $CS_2$ cell: 1.0 mm and 10 mm. One can observe the asymmetry in these time profile curves, due to the relaxation processes in liquid $CS_2$. The symmetric plot (red, solid line) of Figure 5 shows the time profile for the optical gate with 1.0 mm thick gadolinium gallium garnet (GGG) crystal as the Kerr medium. The gate duration (full width at half maximum) for 1.0 mm of this crystal (GGG) is about 0.76 ps whereas for the same thickness of $CS_2$ it is 1.43 ps. GGG has been shown to have high second order refractive index and hence is more efficient for applications in this domain [10]. A more detailed description of this configuration will be presented in the conference including achievable spatial resolution and some preliminary results showing its usability for ballistic imaging of liquid jets/sprays.

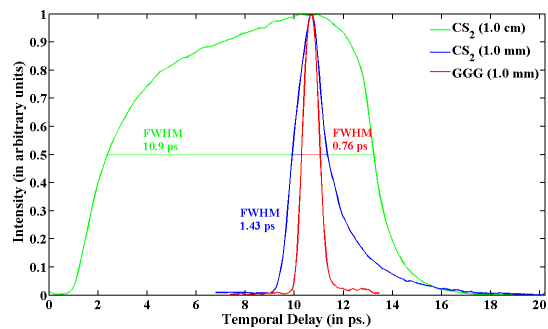

*Figure 5* Temporal profiles of $CS_2$ (1.0 cm and 1.0 mm) and GGG (1.0 mm) in dual colour, collinear configuration of the optical Kerr time gating.

## 4 Conclusions

We have presented a collinear, dual colour Kerr effect based optical time-gate, with GGG and $CS_2$ as Kerr media. Major drawbacks of the classical non-collinear, single colour configuration for the same are discussed and have been resolved with this new setup. The group velocity difference between pump and probe beams inside the Kerr medium does not affect the duration of the gate as long as the thickness of the chosen Kerr medium is small. The temporal characteristics of the optical gate in this new configuration have been presented while spatial resolution estimated using slanted edge method and some preliminary results for the ballistic imaging of liquid sprays/jets will be presented directly at the conference.